\documentclass[english,reprint,aps,amsmath,amssymb,prl,superscriptaddress,floatfix]{revtex4-2}
\usepackage{amsmath}
\usepackage{amssymb}
\usepackage{fontspec}

\usepackage{bm}
\usepackage{graphicx}

\makeatletter

\usepackage{dcolumn}
\usepackage{bm}
\usepackage{mathrsfs}
\usepackage[colorlinks,linkcolor=red,anchorcolor=blue,citecolor=green]{hyperref}
\allowdisplaybreaks[4]
\usepackage{amssymb}
\makeatletter

\newcommand{\Rmnum}[1]{\expandafter\@slowromancap\romannumeral #1@}
\makeatother

\makeatother

\usepackage{polyglossia}
\setdefaultlanguage[variant=american]{english}
\begin{document}
\title{Circular phonon dichroism in $d$-wave altermagnets}
\author{Ding Li}
\affiliation{Anhui Key Laboratory of Low-Energy Quantum Materials and Devices,
High Magnetic Field Laboratory, HFIPS, Chinese Academy of Sciences,
Hefei, Anhui 230031, China}
\affiliation{Department of Physics, University of Science and Technology of China,
Hefei 230026, P.R. China}
\author{Tao Qin}
\affiliation{School of Physics and Optoelectronics Engineering, Anhui University,
Hefei, Anhui Province 230601, P.R. China}
\author{Jianhui Zhou}
\email{jhzhou@hmfl.ac.cn}

\affiliation{Anhui Key Laboratory of Low-Energy Quantum Materials and Devices,
High Magnetic Field Laboratory, HFIPS, Chinese Academy of Sciences,
Hefei, Anhui 230031, China}
\date{\today}
\begin{abstract}
Altermagnets, a new class of collinear antiferromagnets, exhibit momentum-dependent
spin splitting and offer compelling advantages for antiferromagnetic
spintronics. However, the magnetic order is intrinsically difficult
to read out, which hinders practical applications. We propose finite-momentum
circular phonon dichroism as a direct probe that distinguishes the
two opposite out-of-plane Néel vector configurations in two-dimensional
$d$-wave altermagnets. Combining Onsager reciprocity with $C_{2z}$
lattice symmetry, we find that the dichroic signal reverses sign when
the Néel vector is flipped for the in-plane phonon wave vectors. Moreover,
a channel-resolved decomposition identifies the circular phonon dichroism
originates from the interband coherent transitions. Representative
finite-momentum cuts show pronounced dichroic asymmetric ratio, with
$|\eta_{\mathrm{CPD}}|=36\%$. Our work reveals that the circular-ultrasound
absorption can distinguish the two opposite out-of-plane Néel vector
configurations of $d$-wave altermagnets.
\end{abstract}
\maketitle
\textit{Introduction.-{}-}Antiferromagnets offer several advantages
over ferromagnets for information storage: absence of stray fields,
terahertz-scale spin dynamics, and vanishing net magnetization \citep{2018RMP-AFM,2016NatNano-AFM,2018NP-Topological,2018NP-AFMopto}.
Altermagnets, new class of collinear antiferromagnets with momentum-dependent
spin splitting \citep{2026PRB-WuCJ,2024Nature-alter,2022PRX-beyond,2022PRB-Pseudoscalar,2024AFM},
have attracted significant attention not only for their remarkable
functionalities, including the electrical spin-splitter effect, giant
and tunneling magnetoresistance, and spin splitting torque \citep{2021PRL-Efficient,2022PRX-Giant,2025NC-Electrical,2023Nature-Magnetoresistance},
but also as an ideal platform for exploring many novel physical effects
\citep{2024plaid-like,2026arxivAHValter,2026PRL-CW,2025PRL-XXC,2026PRL-CCL,2026PRB-YZB,2023PRL-spin2charge,2024PRX-Octupoles,2026prl-Magnon,2026PRX-SuperSpinCurrents,2024NC-Edelstein,2026PRL-Nernst,2025ShuiD,2025NM-Hall,2026PRL-Proximity,2024NC-SBZ}.
However, determining the orientation of the Néel vector remains a
major challenge \citep{2016Science-Electrical,2024Science-180,2018NC-DetNeel,2017NPho-neel,2026RuiCX,2025Probing}.
Since the spin splitting and associated electronic wave functions
depend sensitively on the Néel order, reversing the Néel vector modifies
the underlying electronic structure and its quantum-geometric properties
\citep{2026PRL-Photocurrent,2024-QGTAlter,2026PRL-HDZ,2026CPL}, providing
a powerful route for detecting the Néel vector of altermagnetic materials.

Circularly polarized phonons can carry angular momentum and chirality
\citep{2014PRL-Angular,2025NP-ChiPhonon,2022PRR-TTZ}, and phonon
dichroisms have been proposed as finite-momentum probes of electronic
dissipation in several quantum materials \citep{2017PRL-Circular,2022PRB-Anomalous}.
It has been shown that both the linear and circular phonon dichroisms
can reveal the electronic quantum geometry through electron–phonon
coupling \citep{2025arxiv-phonon}. These findings naturally suggest
that phonon dichroisms may also encode information about the Néel
vector orientation through its influence on the electronic quantum
geometry \citep{2010RMP-niuqian,1984Berry,2025PRB-MCDAlter,2024PRB-EzawaNon}.
Moreover, phonons intrinsically carry finite momentum, enabling momentum-resolved
probes of electronic dissipation beyond the long-wavelength regime,
in particular, the anisotropic spin splitting Fermi surface in altermagnets.

In this work, we develop a theory of finite-momentum phonon absorption
and demonstrate that circular phonon dichroism (CPD) provides a direct
signature of two opposite out-of-plane Néel vector domains of metallic
$d$-wave altermagnets. Combining Onsager reciprocity with $C_{2z}$
symmetry, we find that the dichroic signal reverses sign when the
Néel vector is flipped for the in-plane phonon wave vectors. A channel-resolved
decomposition identifies the interband-coherence origin of the CPD.
Numerical calculations of $d$-wave altermagnets reveal the strong
modulation of CPD near the resonant finite wave vector. 
\begin{figure}[h]
\centering
\includegraphics[width=8cm]{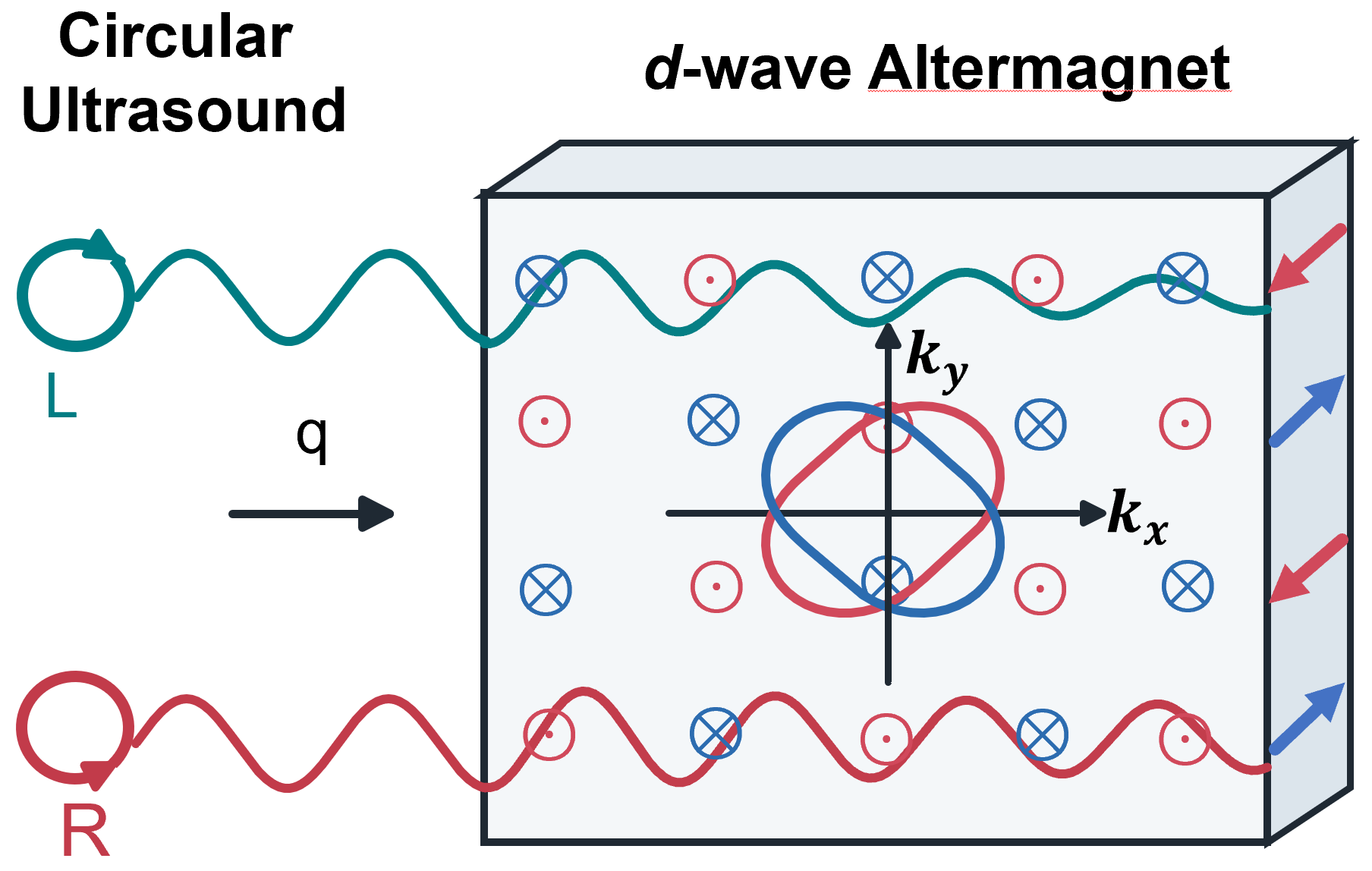}\caption{\label{fig:figure-1-reading-out-the-n-el-vector-ori}Dichroic attenuation
of circular ultrasound through an altermagnetic material. The red
and blue circles denote the local magnetic moments, which are perpendicular
to the page and antiparallel to one another. Equal-amplitude left-
and right-circular ultrasound are transmitted separately through a
$d$-wave altermagnet with a static alternating-spin lattice. The
red and blue contours schematically depict the $d$-wave altermagnetic
Fermi-surface splitting, which vanishes on the crystalline axes and
changes sign between adjacent quadrants. The unequal transmitted amplitudes
correspond to $\gamma^{L}\protect\ne\gamma^{R}$. At fixed and calibrated
$(\boldsymbol{q},\omega)$, the sign of the attenuation difference
distinguishes opposite Néel vector orientations.}
\end{figure}

\textit{Formalism of phonon dichroisms.-{}-}The total Hamiltonian
contains ionic, electronic, and electron–phonon parts, $H_{\mathrm{tot}}=H_{\mathrm{ion}}+H_{\mathrm{el}}+H_{\mathrm{e\text{-}ph}}$
\citep{2009Manybody}. To linear order in the ionic displacement $\boldsymbol{u}(\boldsymbol{q})$,
the electron–phonon coupling is $H_{\mathrm{e\text{-}ph}}=-\sum_{\boldsymbol{q},\alpha}\hat{T}_{\alpha}(-\boldsymbol{q})u_{\alpha}(\boldsymbol{q})$,
where $\hat{T}_{\alpha}(\boldsymbol{q})$ is the generalized force
operator. For simplicity, we consider the lattice dynamics of the
monatomic lattice with an atomic mass $M$ and have the ionic equation
of motion

\begin{equation}
M\ddot{u}_{\alpha}(\boldsymbol{q})=-M\sum_{\beta}\Phi_{\alpha\beta}(\boldsymbol{q})u_{\beta}(\boldsymbol{q})+\langle\hat{T}_{\alpha}(\boldsymbol{q},t)\rangle,
\end{equation}
where $M$ is the ion mass. The electronic feedback force can be calculated
within the linear response regime \citep{2000Mahan} $\langle\hat{T}_{\alpha}(\boldsymbol{q},\omega)\rangle_{\mathrm{ind}}=\hbar\sum_{\beta}\chi_{\alpha\beta}(\boldsymbol{q},\omega)u_{\beta}(\boldsymbol{q},\omega)$,
with the retarded force-force response function

\begin{equation}
\chi_{\alpha\beta}(\boldsymbol{q},\omega)=-\frac{i}{M\hbar}\int^{\infty}_{0}dt\,e^{i(\omega+i0^{+})t}\langle[\hat{T}_{\alpha}(\boldsymbol{q},t),\hat{T}_{\beta}(-\boldsymbol{q},0)]\rangle_{0}.
\end{equation}
where $0^{+}$ is a positive infinitesimal, and the subscript 0 denotes
the expectation value in the unperturbed electronic ground state.
Combining these gives the eigenvalue equation of phonons

\begin{equation}
\omega^{2}u_{\alpha}(\boldsymbol{q})=\sum_{\beta}\left[\hbar^{2}\Phi_{\alpha\beta}(\boldsymbol{q})+\hbar\chi_{\alpha\beta}(\boldsymbol{q},\omega)\right]u_{\beta}(\boldsymbol{q}).
\end{equation}
The first term is the bare ion–ion dynamical matrix in the adiabatic
limit; the second term encodes the quantum geometry informations from
electronic states through the electron-phonon interaction. The dissipative
part of the response function $\chi_{\alpha\beta}$ corresponds to
the anti-Hermitian part: $\gamma_{\alpha\beta}=\frac{i}{4\omega}(\chi_{\alpha\beta}-\chi^{*}_{\beta\alpha})$
and describes the absorptions of phonons such as in Weyl semimetals
and two-dimensional (2D) magnetic Dirac materials \citep{2017PRL-Circular,2022PRB-Anomalous,2023PRR-Nonreciprocal}.
Meanwhile its Hermitian part is associated with the molecular Berry
curvature for phonons that plays a key role in the chiral phonons
and phonon anomalous transport \citep{QinT2012PRB,ZhangLF2015PRL,2022PRR-TTZ,ChenYR2025PRL,ZhangS2026PRB}.
For a phonon polarization $\boldsymbol{\epsilon}$, the absorption
coefficient is $\gamma^{\boldsymbol{\epsilon}}=\boldsymbol{\epsilon}^{\dagger}\gamma\boldsymbol{\epsilon}$.
Decomposing $\gamma$ into symmetric and antisymmetric parts yields

\begin{equation}
\gamma=\begin{pmatrix}\gamma_{D}+\gamma_{\bar{D}} & \gamma^{\bar{A}}+i\gamma^{A}\\
\gamma^{\bar{A}}-i\gamma^{A} & \gamma_{D}-\gamma_{\bar{D}}
\end{pmatrix},
\end{equation}
where $\gamma_{D}=(\gamma_{xx}+\gamma_{yy})/2$ governs polarization-averaged
absorption, $\gamma_{\bar{D}}$ and $\gamma^{\bar{A}}$ govern linear
dichroism, and the Hall-like component $\gamma^{A}=\mathrm{Im}\,\gamma_{xy}$
governs circular dichroism. Thus, a finite CPD requires that $\mathcal{T}$
and $\mathcal{PT}$ symmetry must be broken. The explicit formulas
for the four dissipative components are given in Supplemental Material
(SM) \citep{SI}. For normalized circular polarization vectors $\boldsymbol{\epsilon}^{L/R}=(1,\pm i)/\sqrt{2}$,
one finds $\gamma^{L/R}=\gamma_{D}\mp\gamma^{A}$.

In order to evaluate $\gamma^{A}$ explicitly, we need the force operator
matrix elements at finite phonon momentum. Strain couples to the electronic
momentum through the local frame field \citep{2012PRL-space,2013PRB-Gauge,2009PRL-strain,2007PRB-Symmetry},
in which the ionic displacement field $\boldsymbol{u}(\boldsymbol{r})$
enters through the local frame $e^{\mu}_{j}=\delta^{\mu}_{j}-\partial u_{j}/\partial r_{\mu}$
in the long-wavelength limit, and the electron-phonon coupling follows
from the Hermitian replacement $\hat{p}_{j}\to\tfrac{1}{2}(\hat{p}_{\mu}e^{\mu}_{j}+e^{\mu}_{j}\hat{p}_{\mu})$
the effective Hamiltonian of Bloch electrons in momentum space $h(\boldsymbol{k})$,
which can be usually obtained from the symmetry analysis and first-principles
calculations. For a phonon with wave vector $\boldsymbol{q}$ and
displacement along $\alpha$, the matrix elements of the force operator
are (see details in SM \citep{SI})

\begin{equation}
T^{ss'}_{\alpha}(\boldsymbol{k},\boldsymbol{q})=i\,\boldsymbol{q}\cdot\left(\boldsymbol{k}+\frac{\boldsymbol{q}}{2}\right)\bar{v}_{ss',\alpha}(\boldsymbol{k},\boldsymbol{q}),
\end{equation}
where $\bar{v}_{ss',\alpha}(\boldsymbol{k},\boldsymbol{q})=\left\langle u_{s,\boldsymbol{k}}\left|\frac{\hat{v}_{\alpha}(\boldsymbol{k})+\hat{v}_{\alpha}(\boldsymbol{k}+\boldsymbol{q})}{2}\right|u_{s',\boldsymbol{k}+\boldsymbol{q}}\right\rangle $
is the finite-$\bm{q}$ averaged velocity matrix element.

The dissipative part is then

\begin{equation}
\gamma^{A}(\boldsymbol{q},\omega)=-\frac{1}{2\omega}\sum_{ss'}\int\frac{d^{2}\boldsymbol{k}}{\rho(2\pi)^{2}}F_{s's}(\boldsymbol{k},\boldsymbol{q})\mathrm{Im}[S_{ss'}(\boldsymbol{k},\boldsymbol{q})]
\end{equation}
with $F_{s's}=-\pi[f_{s}(\boldsymbol{k})-f_{s'}(\boldsymbol{k}+\boldsymbol{q})]\delta[\varepsilon_{s'}(\boldsymbol{k}+\boldsymbol{q})-\varepsilon_{s}(\boldsymbol{k})-\hbar\omega]$
and $S_{ss'}(\boldsymbol{k},\boldsymbol{q})=\mathrm{Im}[\bar{v}_{ss',x}\bar{v}^{*}_{ss',y}]\left[\boldsymbol{q}\cdot\left(\boldsymbol{k}+\frac{\boldsymbol{q}}{2}\right)\right]^{2}$.
Here $f_{s}(\boldsymbol{k})$ is the Fermi-Dirac occupation and $\varepsilon_{s}(\boldsymbol{k})$
is the band energy and the typical mass density $\rho=2.2\times10^{-6}\mathrm{kg}/\mathrm{m}^{2}$
for $\mathrm{Ru}\mathrm{O}_{2}$ is chosen in this work. The factor
$\mathrm{Im}F_{s's}$ enforces energy conservation and ensures that
the transition occurs between occupied and empty states.

The essential relation for Néel vector readout follows from time reversal
and Onsager reciprocity \citep{1966PR-SYM}. Under the time-reversal
operation $\mathcal{T}$, the Néel order parameter reverses: the real-space
staggered exchange field $\sum_{i}J\delta(\boldsymbol{r}-\boldsymbol{r}_{i})\sigma_{z}$
on the magnetic sublattice flips through $\sigma_{z}\to-\sigma_{z}$
at fixed spatial profile, equivalently $J\to-J$, the phonon momentum
changes sign ($\boldsymbol{q}\to-\boldsymbol{q}$), and the dissipative
tensor is transposed: $\gamma_{\alpha\beta}(\boldsymbol{q},\omega;J)=\gamma_{\beta\alpha}(-\boldsymbol{q},\omega;-J)$.
Taking the imaginary part of the off-diagonal element, $\gamma^{A}=\mathrm{Im}\,\gamma_{xy}$,
gives rise to the Onsager relation

\begin{equation}
\gamma^{A}(\boldsymbol{q},\omega;J)=-\gamma^{A}(-\boldsymbol{q},\omega;-J).
\end{equation}

If the crystal symmetry maps $-\boldsymbol{q}$ back to $\boldsymbol{q}$
while preserving the scalar quantity $\gamma^{A}$, then $\gamma^{A}(\boldsymbol{q};J)=-\gamma^{A}(\boldsymbol{q};-J)$,
so the CPD changes sign under Néel vector reversal and gives a direct
readout of the Néel orientation. Whether this reduction holds depends
on the crystal point group. We note that the required momentum mapping
is guaranteed for 2D $d$-wave altermagnets.

We note that CPD can be understood by analogy with photon circular
dichroism, such as that probed by X-ray magnetic circular dichroism
(XMCD) \citep{2024PRB-PengR,2025PRBMOC}. Specifically, CPD originates
from the imaginary part of the phonon-mediated response function $\mathrm{Im}\,\gamma_{xy}$,
whereas photon circular dichroism corresponds to the antisymmetric
part of the dynamical anomalous Hall conductivity $\mathrm{Im}\,\sigma_{xy}$
\citep{2020MCD}. Furthermore, a finite CPD requires the electron-phonon
coupling to capture both the stretch and the rotation of the crystal
lattice. The lattice rotation directly generates chiral phonon modes;
the strain $\boldsymbol{u}$ induced by these modes couples to electrons
through the elastic gauge vector potential $\hat{\boldsymbol{A}}(\boldsymbol{u})$.
The resulting electron-phonon coupling has the form $\hat{H}_{\mathrm{e-ph}}=g\hat{\boldsymbol{v}}\cdot\hat{\boldsymbol{A}}(\boldsymbol{u})$,
where g is the coupling constant and $\hat{\boldsymbol{v}}$ is the
electron velocity operator that induces interband hopping of electrons.

\begin{figure}
\includegraphics[width=8cm]{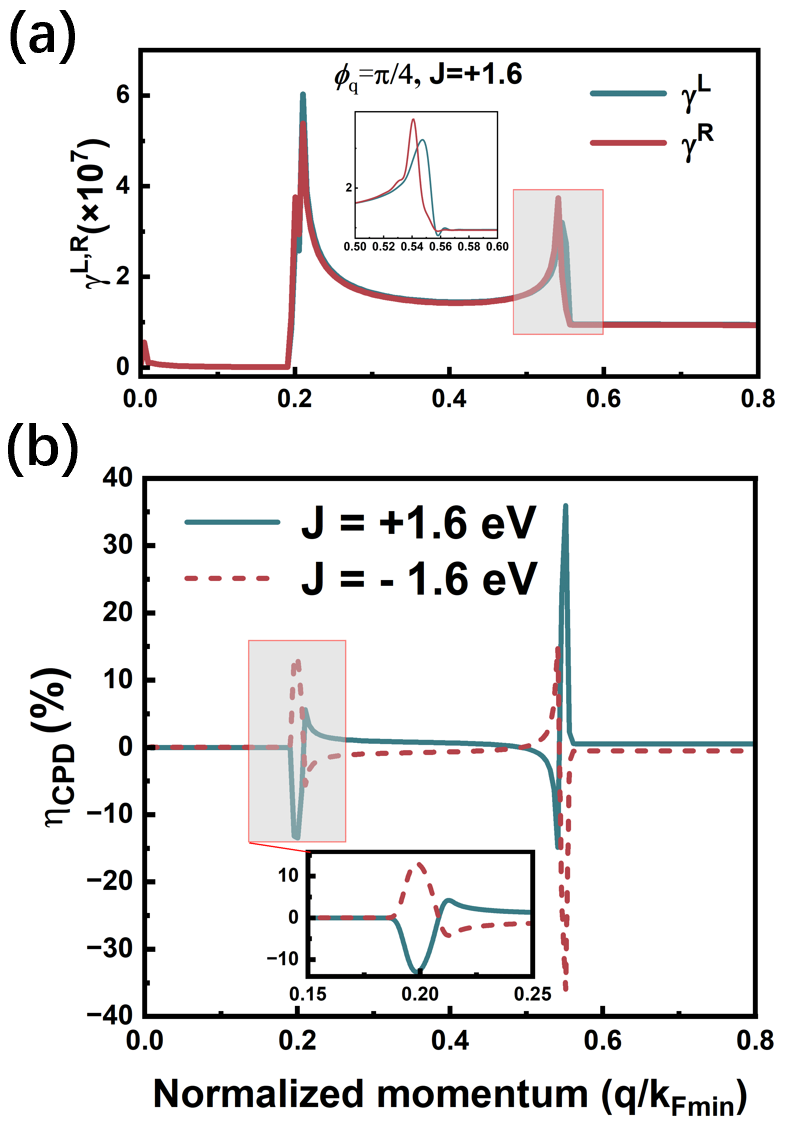}\caption{\label{fig:figure-2-finite-momentum-cpd-scan-at-a-f}Radial circular-channel
absorption and CPD asymmetry ratio. (a) Momentum dependence of $\gamma^{L}$
and $\gamma^{R}$(1/s) along $\phi_{q}=\pi/4$ for $J=1.6\,\mathrm{eV}$.
(b) Normalized CPD asymmetry ratio $\eta_{\mathrm{CPD}}$ along the
same direction for $J=\pm1.6\,\mathrm{eV}$. Solid and dashed curves
denote $J=1.6\,\mathrm{eV}$ and $J=-1.6\,\mathrm{eV}$, respectively.
The inset enlarges the low-$q$ resonance region.}
\end{figure}

\textit{$d$-wave altermagnets.-{}-}We focus on the 2D $d$-wave altermagnet
illustrated in Fig. \ref{fig:figure-1-reading-out-the-n-el-vector-ori},
with the Néel vector along the $z$ axis and phonons propagating in
the $x$-$y$ plane. Representative material platforms include $\mathrm{Ru}\mathrm{O}_{2}$
and $\mathrm{K}\mathrm{V}_{2}\mathrm{Se}_{2}\mathrm{O}$ \citep{2025PRL-YanSun,2024SciAdv-Observation,2020SciA-Crystal,2025NP-TianQ}.
Since the the electron-hole excitations by the low-energy acoustic
phonons (meV scale) get greatly suppressed due to the finite band
gap. Thus, we focus on the metallic case and consider the minimal
two-band Hamiltonian for $d$-wave altermagnets \citep{2024PRB-Minimal,2024NC-SBZ}

\begin{equation}
h(\boldsymbol{k})=tk^{2}\sigma_{0}+Jk_{x}k_{y}\sigma_{z}+\lambda(k_{x}\sigma_{y}-k_{y}\sigma_{x}),
\end{equation}
where $J$ is the $d$-wave altermagnetic exchange splitting \citep{ExchangeJ}and
$\lambda$ is the Rashba spin-orbit coupling \citep{1985Rashba-Spectrum}.
The parameters used here are $\mu=0.4$ eV, $t=2$ eV, $\lambda=0.1$
eV, $J=1.6$ eV, $\hbar\omega=0.002$ eV, and $T=20$ K \citep{2024NC-SBZ,2024PRB-Minimal,2025PRLDrag}.
Reversing the Néel vector flips $\sigma_{z}$, which is equivalent
to $J\to-J$ since the exchange splitting enters only through $Jk_{x}k_{y}\sigma_{z}$.
The Rashba term is essential for the Hall-like response: without it
($\lambda=0$) the Berry curvature vanishes and $\gamma^{A}$ is identically
zero \citep{2010RMP-AHE}. The band dispersion is $\varepsilon_{\pm}(\boldsymbol{k})=tk^{2}\pm\sqrt{J^{2}k^{2}_{x}k^{2}_{y}+\lambda^{2}k^{2}}$.
This $d$-wave exchange splitting vanishes on the crystalline axes
and changes sign between adjacent quadrants, producing the alternating
spin-split Fermi contours shown in Fig. \ref{fig:figure-1-reading-out-the-n-el-vector-ori}.
We note that Rashba SOC is not a general requirement for CPD \citep{RashbaSOC},
which could be induced by other forms of SOC or the magnon-phonon
coupling \citep{Nomura2019PRL,Weibenhofer2025PRL}.

It is known that 2D $d$-wave altermagnet retains $C_{2z}$ symmetry
\citep{2024PRX-zhiDS,2024PRB-CrystalGroup,2024PRX-SSGQHL}, which
maps $\boldsymbol{q}\to-\boldsymbol{q}$ at fixed $J$ while preserving
the $\gamma^{A}$. Combining $C_{2z}$ with the Onsager relation gives

\begin{equation}
\gamma^{A}(\boldsymbol{q},\omega;J)\rightarrow-\gamma^{A}(\boldsymbol{q},\omega;-J).
\end{equation}

This symmetry relation translates into the transmission protocol shown
in Fig. \ref{fig:figure-1-reading-out-the-n-el-vector-ori}. At a
fixed and calibrated $(\boldsymbol{q},\omega)$, the left- and right-circular
ultrasound with equal incident amplitudes are sent through the sample
in separate measurements. Their unequal attenuation gives $\gamma^{L}\ne\gamma^{R}$,
while reversing the Néel vector $\boldsymbol{n}$ changes the sign
of the transmission difference, $\mathrm{\gamma^{A}(}\boldsymbol{n})=-\gamma^{A}(-\boldsymbol{n})$.
Thus the readout is a sign-sensitive distinction between the two opposite
Néel vector configurations rather than an unconstrained reconstruction
of an arbitrary vector direction. Note that the impacts of the generic
direction of the Néel vector on $\mathrm{\gamma^{A}(}\boldsymbol{n})$
could be treated as that in transport properties \citep{SI,2022PRX-Giant,2024PRB-PengR}.

Fig. \ref{fig:figure-2-finite-momentum-cpd-scan-at-a-f}(a) shows
the two circular-channel absorptions $\gamma^{L}$ and $\gamma^{R}$
at different $\boldsymbol{q}$ along $\phi_{q}=\pi/4$ for $J=1.6\,\mathrm{eV}$.
They share the same resonant absorption background but separate at
selected momenta. Their difference is the CPD and their average gives
the polarization-averaged scale $\gamma_{D}$. Since the background
itself varies strongly with $q$, we quantify the effect by the asymmetric
ratio $\eta_{\mathrm{CPD}}$, defined as 
\begin{equation}
\eta_{\mathrm{CPD}}=\frac{\gamma^{L}-\gamma^{R}}{\gamma^{L}+\gamma^{R}}=-\frac{\gamma^{A}}{\gamma_{D}}.
\end{equation}

Fig. \ref{fig:figure-2-finite-momentum-cpd-scan-at-a-f}(b) puts this
ratio on the same radial axis for the two Néel vector orientations,
represented by $J=\pm1.6\,\mathrm{eV}$. The sign reversal between
the two curves is visible throughout the scanned radial range, consistent
with the symmetry relation above. Along this cut, the signal is not
a smooth function of $\boldsymbol{q}$: it exhibits a local peak at
$q_{1}/k_{F,\min}=0.2$ and a second, more pronounced peak with $|\eta_{\mathrm{CPD}}|\simeq36\%$,
at $q_{2}/k_{F,\min}=0.55$, where $k_{F,\min}=\sqrt{\mu/(t+|J|/2)}$.
Consequently, since $\eta_{\mathrm{CPD}}$ changes sign at any fixed
momentum upon Néel vector reversal, measuring at a single momentum
point is sufficient for detecting the switch. 
\begin{figure}
\includegraphics[width=8.5cm]{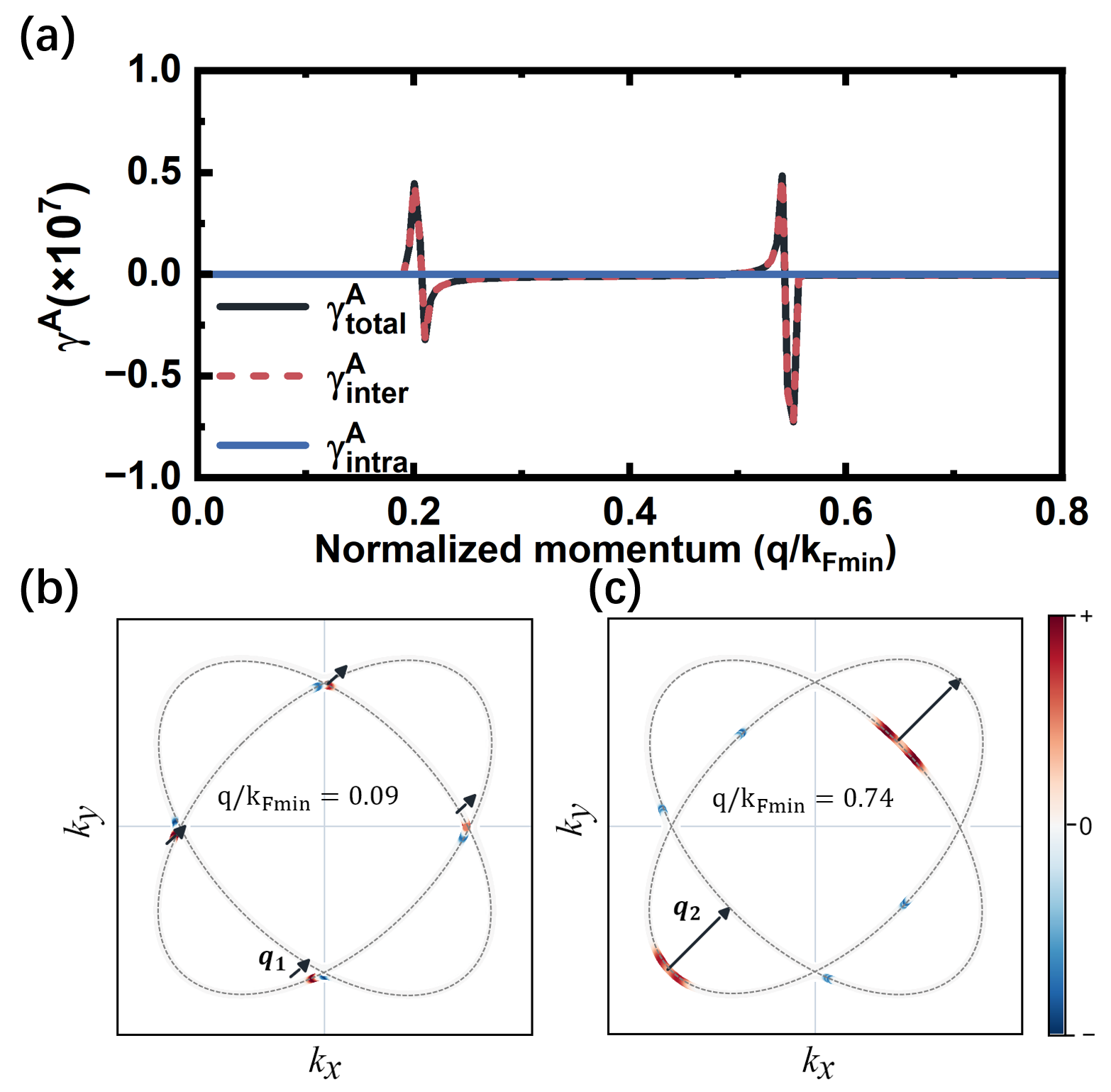}\caption{\label{fig:figure-3-directional-scan-of-the-finite}Channel and $k$-resolved
origin of the radial resonances. (a) Channel-resolved $\gamma^{A}$(1/s)
for $J=1.6\,\mathrm{eV}$: total (solid black), interband (dashed
red), and intraband (solid blue). Interband kernel $g_{A}(\boldsymbol{k},\boldsymbol{q})$
of $\gamma^{A}$ plotted as $\mathrm{sgn}(g_{A})|g_{A}/g_{A,\max}|$
on the SOC Fermi surfaces for (b) $q_{1}/k_{F,\min}=0.2$ and (c)
$q_{2}/k_{F,\min}=0.55$. Red and blue denote positive and negative
values after separate panel normalization. Dashed lines show the SOC
Fermi surfaces. Arrows indicate representative near-Fermi transitions
$\boldsymbol{k}_{i}\to\boldsymbol{k}_{i}+\boldsymbol{q}_{1,2}$ selected
by the corresponding peak momenta.}
\end{figure}

To uncover the microscopic origin of this signal, we decompose $\gamma^{A}$
into band-transition channels for $J=1.6\,\mathrm{eV}$ in Fig. \ref{fig:figure-3-directional-scan-of-the-finite}(a).
The interband contribution $\gamma^{A}_{{\rm inter}}$ is visually
indistinguishable from the total $\gamma^{A}$ across the scan, with
the intraband contribution staying below $10^{-4}$ of the peak. In
this model, the CPD at each $q$ therefore arises from interband coherent
transitions between the spin-splitting branches; the leading intraband
density-like term carries zero cross imaginary part and does not contribute
to the net chiral response. Figs. \ref{fig:figure-3-directional-scan-of-the-finite}(b,
c) show the $\boldsymbol{k}$-resolved kernel of $\gamma^{A}$ at
$q_{1}$ and $q_{2}$. We plot $\mathrm{sgn}(g_{A})|g_{A}/g_{A,\max}|$,
where $g_{A}(\boldsymbol{k},\boldsymbol{q})$ is the interband integrand
of $\gamma^{A}$ including all coefficients and $g_{A,\max}$ is its
maximum absolute value within each panel. The kernel is shown on the
SOC Fermi surface prior to $\boldsymbol{k}$ integration, so the color
encodes sign and relative weight within each panel. For $q_{1}$,
four small regions near the crystalline axes carry the weight. For
$q_{2}$, two larger patches dominate and are connected by the indicated
finite-$q$ transitions. The peaks originate from electron transitions
due to the phonons at multiple regions near the Fermi surface. In
the finite-$q$ response, the energy conservation restricts the integrand
to $\varepsilon_{s'}(\boldsymbol{k}+\boldsymbol{q})-\varepsilon_{s}(\boldsymbol{k})=\hbar\omega$.
When this contour nearly touches a Fermi surface, or when its normal
energy gradient becomes small, the joint density of states increases.
Sweeping $q$ then opens and closes resonant arcs and rearranges the
positive and negative chiral weights. This produces both the peaks
and the intervening zero crossings.

The radial cut establishes a representative signal scale and identifies
a near-resonant momentum shell for angular readout. We therefore fix
$|\boldsymbol{q}|/k_{F,\min}=0.55$, slightly below the second peak
to avoid numerical sensitivity at the exact resonance, and scan the
propagation angle.

\begin{figure}
\includegraphics[width=9cm]{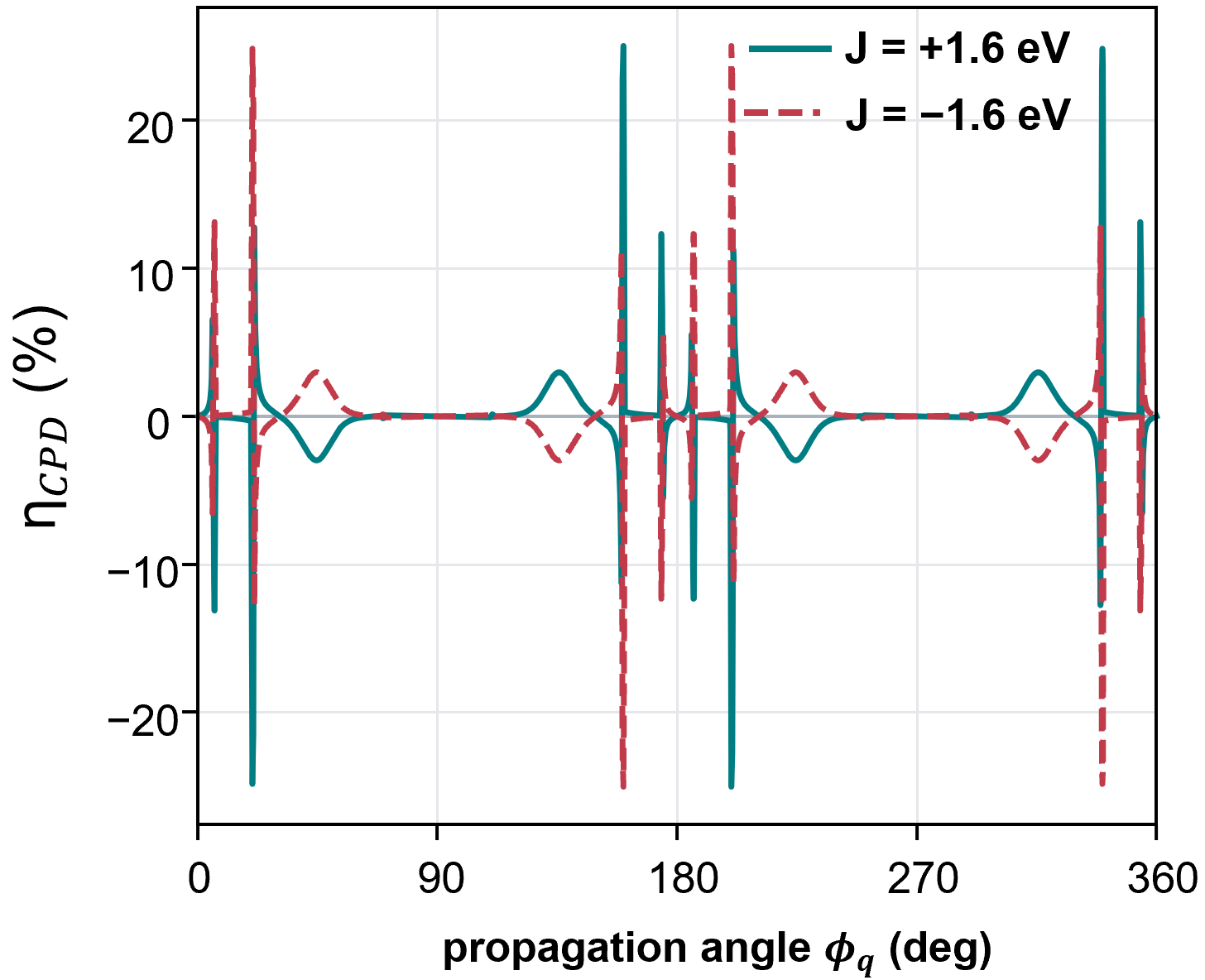}\caption{\label{fig:figure-4-directional-scan-of-the-finite}Directional scan
of the finite-momentum CPD for $J=\pm1.6\,\mathrm{eV}$. The momentum
magnitude is fixed at $|\boldsymbol{q}|/k_{F,\min}=0.55$.}
\end{figure}
 On this momentum shell, theangular scan reaches its largest absolute
asymmetric ratio with $J=1.6\,\mathrm{eV}$, $|\eta_{\mathrm{CPD}}|=25\%$.
Reversing the Néel vector ($J$ to $-J$) reverses the entire angular
curve up to numerical precision. The CPD also vanishes at $\phi_{q}=0^{\circ}$,
$90^{\circ}$, $180^{\circ}$, and $270^{\circ}$, consistent with
the mirror symmetries $M_{x}$ and $M_{y}$, which make the CPD odd
under $q_{x}\to-q_{x}$ and $q_{y}\to-q_{y}$, respectively. Experimentally,
the 2$\pi$-azimuthal scan shown in Fig. \ref{fig:figure-4-directional-scan-of-the-finite}
can be realized with multiple shear-wave transducers placed at different
azimuthal angles around the sample.

Beyond these symmetry-enforced zeros, the CPD changes sign additionally
within each quadrant. These finite-angle lobes provide the practical
readout points: once a resonant momentum shell is selected, the sign
of the circular attenuation asymmetry distinguishes opposite Néel
vector orientations. Thus symmetry fixes the Néel vector sign relation
and the four axis zeros, whereas the heights, and extra sign changes
are controlled by the phonon assisted electron transitions with finite-momentum
transfer. In addition, we also give the results of linear phonon dichroism
in SM \citep{SI}.

We briefly discuss the relation between CPD and the Fermi-surface
geometry and Néel vector. As illustrated in Fig. \ref{fig:figure-3-directional-scan-of-the-finite},
CPD originates from interband transitions between spin-split bands.
The two transition channels, up→down and down→up, contribute to $\eta_{\mathrm{CPD}}$
with opposite signs (see Fig. \ref{fig:figure-4-directional-scan-of-the-finite}),
so the sign and magnitude of $\eta_{\mathrm{CPD}}$ directly reflect
the spin-splitting structure of the Fermi surface. In the $d$-wave
altermagnet considered here, this leads to a characteristic fourfold
angular dependence of $\eta_{\mathrm{CPD}}$, which encodes the $d$-wave
symmetry of the Fermi surface. The same analysis could be applicable
to other altermagnets, such as $f$-wave and $g$-wave, where the
corresponding angular dependence of $\eta_{\mathrm{CPD}}$ would reflect
the specific Fermi-surface geometry \citep{SI}. Furthermore, because
CPD is sensitive to the spin splitting and its sign reverses upon
reversing the Néel vector, it provides a potentially useful probe
of both the Fermi-surface geometry and the magnetic order in altermagnets.

Experimentally, the pulse-echo ultrasound method provides a direct
probe of CPD. The directly measured quantities are the incident amplitude
$A_{0}$ and the two transmitted amplitudes $A^{L}$ and $A^{R}$.
For a sample of thickness $d$ and sound velocity $c_{t}$, the damping
rates satisfy $\gamma^{L/R}=-(\frac{c_{t}}{d})\ln(\frac{A^{L/R}}{A_{0}})$
\citep{2016Ultrasonic}, so 
\begin{equation}
\eta_{\mathrm{CPD}}=-\frac{\ln(A^{L}/A^{R})}{\ln\bigl(A^{2}_{0}/(A^{L}A^{R})\bigr)}
\end{equation}
Thus $\eta_{\mathrm{CPD}}$ is determined entirely by the measured
amplitudes; no knowledge of $d$, $c_{t}$, or other material-specific
parameters is required. For a typical transverse acoustic velocity
$c_{t}\sim3\times10^{3}\mathrm{m}/\mathrm{s}$ and the damping difference
$\gamma^{L}-\gamma^{R}\sim10^{7}/\mathrm{s}$ at $|\boldsymbol{q}|/k_{F,\min}=0.55$,
the attenuation difference is estimated to be on the order of $10^{3}/\mathrm{m}$(roughly
8 $\mathrm{dB}/\mathrm{mm}$), which is well above the detection threshold
of current ultrasonic technology \citep{1961PR-abs}. The two opposite
Néel vector orientations can be distinguished by the sign of the attenuation
difference. We note that in 2D materials the in-plane and out-of-plane
transverse modes are non-degenerate, so an injected circular wave
splits into two linearly polarized components. However, CPD is a local
absorption property and can also be probed by far-infrared optical
spectroscopy without requiring the wave to maintain circular polarization,
as demonstrated in a recent experiment \citep{2025PRLCPD-Exp}. Moreover,
realistic effects such as finite temperature, disorder, and phonon
damping are expected to reduce the peak amplitudes and broaden the
sharp features in $\eta_{\mathrm{CPD}}$. However, the sign reversal
of $\eta_{\mathrm{CPD}}$ upon Néel vector reversal is symmetry-enforced
and therefore remains robust. Thus, even with realistic broadening,
the CPD signal provides an unambiguous signature of Néel vector switching.

\textit{Conclusions.-{}-}We developed a finite-momentum theory of
CPD for $d$-wave altermagnets and revealed its origin of the chiral
interband matrix elements. Meanwhile, the CPD is odd under Néel vector
reversal and shows a strong wave vector modulation near resonant phonon
momenta. Specifically, representative finite-momentum cuts show pronounced
dichroic resonances: the radial cut reaches $|\eta_{\mathrm{CPD}}|=36\%$,
and the selected angular scan reaches $|\eta_{\mathrm{CPD}}|=25\%$.
Thus, our work offers an acoustic probe of the Néel vector of $d$-wave
altermagnets, which can be straightforwardly applicable to other altermagnetic
materials.

This work was financially supported by the National Key R\&D Program
of the MOST of China (Grant No. 2024YFA1611300), the National Natural
Science Foundation of China (Grant No. 12574059), HFIPS Director’s
Fund (Grant No. BJPY2023B05), Anhui Provincial Major S\&T Project
(s202305a12020005) and the Basic Research Program of the Chinese Academy
of Sciences Based on Major Scientific Infrastructures (Grant No. JZHKYPT-2021-08)
and the High Magnetic Field Laboratory of Anhui Province under contract
No. AHHM-FX-2020-02.

\bibliographystyle{apsrev4-2}
\bibliography{alterPD}

@article{ChenYR2025PRL,
  title = {Emergence of Chiral Phonons in Two-Dimensional Kagome Lattices Harboring Electronic Chirality},
  author = {Chen, Yanru and Qin, Wei and Zhang, Shunhong and Cui, Ping and Niu, Qian and Zhang, Zhenyu},
  journal = {Phys. Rev. Lett.},
  volume = {135},
  issue = {12},
  pages = {126608},
  numpages = {9},
  year = {2025},
  month = {Sep},
  publisher = {American Physical Society},
  doi = {10.1103/bfll-sdrb},
  url = {https://link.aps.org/doi/10.1103/bfll-sdrb}
}

@article{QinT2012PRB,
  title = {Berry curvature and the phonon Hall effect},
  author = {Qin, Tao and Zhou, Jianhui and Shi, Junren},
  journal = {Phys. Rev. B},
  volume = {86},
  issue = {10},
  pages = {104305},
  numpages = {9},
  year = {2012},
  month = {Sep},
  publisher = {American Physical Society},
  doi = {10.1103/PhysRevB.86.104305},
  url = {https://link.aps.org/doi/10.1103/PhysRevB.86.104305}
}

@article{ZhangLF2015PRL,
  title = {Chiral Phonons at High-Symmetry Points in Monolayer Hexagonal Lattices},
  author = {Zhang, Lifa and Niu, Qian},
  journal = {Phys. Rev. Lett.},
  volume = {115},
  issue = {11},
  pages = {115502},
  numpages = {5},
  year = {2015},
  month = {Sep},
  publisher = {American Physical Society},
  doi = {10.1103/PhysRevLett.115.115502},
  url = {https://link.aps.org/doi/10.1103/PhysRevLett.115.115502}
}

@article{ZhangS2026PRB,
  title = {Comprehensive study of phonon chirality under symmetry constraints},
  author = {Zhang, Shuai and Huang, Zhiheng and Du, Muchen and Ying, Tianping and Du, Luojun and Zhang, Tiantian},
  journal = {Phys. Rev. B},
  volume = {113},
  issue = {2},
  pages = {024302},
  numpages = {10},
  year = {2026},
  month = {Jan},
  publisher = {American Physical Society},
  doi = {10.1103/gmfc-gx4v},
  url = {https://link.aps.org/doi/10.1103/gmfc-gx4v}
}

@article{Weibenhofer2025PRL,
  title = {Chiral Phonons Arising from Chirality-Selective Magnon-Phonon Coupling},
  author = {Wei\ss{}enhofer, Markus and Rieger, Philipp and Mrudul, M. S. and Mikadze, Luca and Nowak, Ulrich and Oppeneer, Peter M.},
  journal = {Phys. Rev. Lett.},
  volume = {135},
  issue = {21},
  pages = {216701},
  numpages = {8},
  year = {2025},
  month = {Nov},
  publisher = {American Physical Society},
  doi = {10.1103/j7bs-2zbx},
  url = {https://link.aps.org/doi/10.1103/j7bs-2zbx}
}

@article{Nomura2019PRL,
  title = {Phonon Magnetochiral Effect},
  author = {Nomura, T. and Zhang, X.-X. and Zherlitsyn, S. and Wosnitza, J. and Tokura, Y. and Nagaosa, N. and Seki, S.},
  journal = {Phys. Rev. Lett.},
  volume = {122},
  issue = {14},
  pages = {145901},
  numpages = {5},
  year = {2019},
  month = {Apr},
  publisher = {American Physical Society},
  doi = {10.1103/PhysRevLett.122.145901},
  url = {https://link.aps.org/doi/10.1103/PhysRevLett.122.145901}
}

@article{2018RMP-AFM,
  title = {Antiferromagnetic Spintronics},
  author = {Baltz, V. and Manchon, A. and Tsoi, M. and Moriyama, T. and Ono, T. and Tserkovnyak, Y.},
  year = 2018,
  month = feb,
  journal = {Reviews of Modern Physics},
  volume = {90},
  number = {1},
  pages = {015005},
  publisher = {American Physical Society},
  doi = {10.1103/RevModPhys.90.015005},
  urldate = {2026-06-24}
}

@article{2016NatNano-AFM,
  title = {Antiferromagnetic Spintronics},
  author = {Jungwirth, T. and Marti, X. and Wadley, P. and Wunderlich, J.},
  year = 2016,
  month = mar,
  journal = {Nature Nanotechnology},
  volume = {11},
  number = {3},
  pages = {231--241},
  publisher = {Nature Publishing Group},
  issn = {1748-3395},
  doi = {10.1038/nnano.2016.18},
  urldate = {2026-06-24},
  copyright = {2016 Springer Nature Limited},
  langid = {english}
}

@article{2018NP-Topological,
  title = {Topological Antiferromagnetic Spintronics},
  author = {{\v S}mejkal, Libor and Mokrousov, Yuriy and Yan, Binghai and MacDonald, Allan H.},
  year = 2018,
  month = mar,
  journal = {Nature Physics},
  volume = {14},
  number = {3},
  pages = {242--251},
  publisher = {Nature Publishing Group},
  issn = {1745-2481},
  doi = {10.1038/s41567-018-0064-5},
  urldate = {2026-06-24},
  copyright = {2018 Springer Nature Limited},
  langid = {english}
  }

@article{2018NP-AFMopto,
  title = {Antiferromagnetic Opto-Spintronics},
  author = {N{\v e}mec, P. and Fiebig, M. and Kampfrath, T. and Kimel, A. V.},
  year = 2018,
  month = mar,
  journal = {Nature Physics},
  volume = {14},
  number = {3},
  pages = {229--241},
  publisher = {Nature Publishing Group},
  issn = {1745-2481},
  doi = {10.1038/s41567-018-0051-x},
  urldate = {2026-06-24},
  copyright = {2018 Springer Nature Limited},
  langid = {english}
  }

@article{2026PRB-WuCJ,
  title = {Unconventional Magnetism in Spin-Orbit Coupled Systems},
  author = {Yuan, Jian-Keng and Pan, Zhiming and Wu, Congjun},
  year = 2026,
  month = jan,
  journal = {Physical Review B},
  volume = {113},
  number = {1},
  pages = {014426},
  publisher = {American Physical Society},
  doi = {10.1103/dy2j-mc2t},
  urldate = {2026-06-30},
  langid = {english}
}

@article{2024Nature-alter,
  title = {Altermagnetic Lifting of {{Kramers}} Spin Degeneracy},
  author = {Krempask{\'y}, J. and {\v S}mejkal, L. and D'Souza, S. W. and Hajlaoui, M. and Springholz, G. and Uhl{\'i}{\v r}ov{\'a}, K. and Alarab, F. and Constantinou, P. C. and Strocov, V. and Usanov, D. and Pudelko, W. R. and {Gonz{\'a}lez-Hern{\'a}ndez}, R. and Birk Hellenes, A. and Jansa, Z. and Reichlov{\'a}, H. and {\v S}ob{\'a}{\v n}, Z. and Gonzalez Betancourt, R. D. and Wadley, P. and Sinova, J. and Kriegner, D. and Min{\'a}r, J. and Dil, J. H. and Jungwirth, T.},
  year = 2024,
  month = feb,
  journal = {Nature},
  volume = {626},
  number = {7999},
  pages = {517--522},
  publisher = {Nature Publishing Group},
  issn = {1476-4687},
  doi = {10.1038/s41586-023-06907-7},
  urldate = {2026-06-24},
  copyright = {2024 The Author(s)},
  langid = {english}
  }

@article{2022PRX-beyond,
  title = {Beyond {{Conventional Ferromagnetism}} and {{Antiferromagnetism}}: {{A Phase}} with {{Nonrelativistic Spin}} and {{Crystal Rotation Symmetry}}},
  shorttitle = {Beyond {{Conventional Ferromagnetism}} and {{Antiferromagnetism}}},
  author = {{\v S}mejkal, Libor and Sinova, Jairo and Jungwirth, Tomas},
  year = 2022,
  month = sep,
  journal = {Physical Review X},
  volume = {12},
  number = {3},
  pages = {031042},
  publisher = {American Physical Society},
  doi = {10.1103/PhysRevX.12.031042},
  urldate = {2026-06-24},
  langid = {english}
  }

@article{2022PRB-Pseudoscalar,
  title = {Altermagnetism and Magnetic Groups with Pseudoscalar Electron Spin},
  author = {Turek, Ilja},
  year = 2022,
  month = sep,
  journal = {Physical Review B},
  volume = {106},
  number = {9},
  pages = {094432},
  issn = {2469-9950, 2469-9969},
  doi = {10.1103/PhysRevB.106.094432},
  urldate = {2023-07-19},
  langid = {english},
}

@article{2024AFM,
author = {Bai, Ling and Feng, Wanxiang and Liu, Siyuan and Šmejkal, Libor and Mokrousov, Yuriy and Yao, Yugui},
title = {Altermagnetism: Exploring New Frontiers in Magnetism and Spintronics},
journal = {Advanced Functional Materials},
volume = {34},
number = {49},
pages = {2409327},
doi = {https://doi.org/10.1002/adfm.202409327},
year = {2024}
}

@article{2021PRL-Efficient,
  title = {Efficient {{Electrical Spin Splitter Based}} on {{Nonrelativistic Collinear Antiferromagnetism}}},
  author = {{Gonz{\'a}lez-Hern{\'a}ndez}, Rafael and {\v S}mejkal, Libor and V{\'y}born{\'y}, Karel and Yahagi, Yuta and Sinova, Jairo and Jungwirth, Tom{\'a}{\v s} and {\v Z}elezn{\'y}, Jakub},
  year = 2021,
  month = mar,
  journal = {Physical Review Letters},
  volume = {126},
  number = {12},
  pages = {127701},
  publisher = {American Physical Society},
  doi = {10.1103/PhysRevLett.126.127701},
  urldate = {2026-06-24},
  langid = {english}
}

@article{2022PRX-Giant,
  title = {Giant and {{Tunneling Magnetoresistance}} in {{Unconventional Collinear Antiferromagnets}} with {{Nonrelativistic Spin-Momentum Coupling}}},
  author = {{\v S}mejkal, Libor and Hellenes, Anna Birk and {Gonz{\'a}lez-Hern{\'a}ndez}, Rafael and Sinova, Jairo and Jungwirth, Tomas},
  year = 2022,
  month = feb,
  journal = {Physical Review X},
  volume = {12},
  number = {1},
  pages = {011028},
  publisher = {American Physical Society},
  doi = {10.1103/PhysRevX.12.011028},
  urldate = {2026-06-24},
  langid = {english}
  }

@article{2025NC-Electrical,
  title = {Electrical Manipulation of Spin Splitting Torque in Altermagnetic {{RuO2}}},
  author = {Zhang, Yichi and Bai, Hua and Dai, Jiankun and Han, Lei and Chen, Chong and Liang, Shixuan and Cao, Yanzhang and Zhang, Yingying and Wang, Qian and Zhu, Wenxuan and Pan, Feng and Song, Cheng},
  year = 2025,
  month = jul,
  journal = {Nature Communications},
  volume = {16},
  number = {1},
  pages = {5646},
  publisher = {Nature Publishing Group},
  issn = {2041-1723},
  doi = {10.1038/s41467-025-60891-2},
  urldate = {2026-06-24},
  copyright = {2025 The Author(s)},
  langid = {english}
  }

@article{2023Nature-Magnetoresistance,
  title = {Room-Temperature Magnetoresistance in an All-Antiferromagnetic Tunnel Junction},
  author = {Qin, Peixin and Yan, Han and Wang, Xiaoning and Chen, Hongyu and Meng, Ziang and Dong, Jianting and Zhu, Meng and Cai, Jialin and Feng, Zexin and Zhou, Xiaorong and Liu, Li and Zhang, Tianli and Zeng, Zhongming and Zhang, Jia and Jiang, Chengbao and Liu, Zhiqi},
  year = 2023,
  month = jan,
  journal = {Nature},
  volume = {613},
  number = {7944},
  pages = {485--489},
  publisher = {Nature Publishing Group},
  issn = {1476-4687},
  doi = {10.1038/s41586-022-05461-y},
  urldate = {2026-06-30},
  copyright = {2023 The Author(s), under exclusive licence to Springer Nature Limited},
  langid = {english}
}

@article{2024plaid-like,
  author = {Zhu, Yu-Peng and Chen, Xiaobing and Liu, Xiang-Rui and Liu, Yuntian and Liu, Pengfei and Zha, Heming and Qu, Gexing and Hong, Caiyun and Li, Jiayu and Jiang, Zhicheng and Ma, Xiao-Ming and Hao, Yu-Jie and Zhu, Ming-Yuan and Liu, Wenjing and Zeng, Meng and Jayaram, Sreehari and Lenger, Malik and Ding, Jianyang and Mo, Shu and Tanaka, Kiyohisa and Arita, Masashi and Liu, Zhengtai and Ye, Mao and Shen, Dawei and Wrachtrup, J{\"o}rg and Huang, Yaobo and He, Rui-Hua and Qiao, Shan and Liu, Qihang and Liu, Chang},
  title = {Observation of plaid-like spin splitting in a noncoplanar antiferromagnet},
  journal = {Nature},
  year = {2024},
  volume = {626},
  number = {7999},
  pages = {523--528},
  doi = {10.1038/s41586-024-07023-w},
  url = {https://doi.org/10.1038/s41586-024-07023-w}
}

@article{2024NC-Edelstein,
  title = {Non-Relativistic Torque and {{Edelstein}} Effect in Non-Collinear Magnets},
  author = {{Gonz{\'a}lez-Hern{\'a}ndez}, Rafael and Ritzinger, Philipp and V{\'y}born{\'y}, Karel and {\v Z}elezn{\'y}, Jakub and Manchon, Aur{\'e}lien},
  year = 2024,
  month = sep,
  journal = {Nature Communications},
  volume = {15},
  number = {1},
  pages = {7663},
  issn = {2041-1723},
  doi = {10.1038/s41467-024-51565-6},
  urldate = {2025-12-09},
  langid = {english},
}

@article{2026prl-Magnon,
  title = {Interaction-Driven Altermagnetic Magnon Chiral Splitting},
  author = {Jin, Zhejunyu and Zeng, Zhaozhuo and Liu, Jie and Gong, Tianci and Su, Ying and Chang, Kai and Yan, Peng},
  journal = {Phys. Rev. Lett.},
  volume = {136},
  issue = {8},
  pages = {086703},
  numpages = {9},
  year = {2026},
  month = {Feb},
  publisher = {American Physical Society},
  doi = {10.1103/v867-h742},
  url = {https://link.aps.org/doi/10.1103/v867-h742}
}

@article{2026PRL-Nernst,
  title = {Altermagnet-Driven Magnon Spin Splitting Nernst Effect},
  author = {Yang, Yuben and Wang, Di and Yang, Bin and Wang, Peng and Mu, Yuxuan and Tian, Yuanzhe and Zheng, Bowen and Qin, Weijie and Wang, Kaiyuan and Huang, Biying and Wang, Baigeng and Wan, Xiangang and Wu, Di},
  journal = {Phys. Rev. Lett.},
  volume = {136},
  issue = {2},
  pages = {026701},
  numpages = {7},
  year = {2026},
  month = {Jan},
  publisher = {American Physical Society},
  doi = {10.1103/g5xq-z15c},
  url = {https://link.aps.org/doi/10.1103/g5xq-z15c}
}

@article{2026PRX-SuperSpinCurrents,
  title = {Persistent Spin Currents in Superconducting Altermagnets},
  author = {Monkman, Kyle and Weng, Joan and Heinsdorf, Niclas and Nocera, Alberto and Barlas, Yafis and Franz, Marcel},
  journal = {Phys. Rev. X},
  volume = {16},
  issue = {1},
  pages = {011057},
  numpages = {21},
  year = {2026},
  month = {Mar},
  publisher = {American Physical Society},
  doi = {10.1103/52wh-1z5y},
  url = {https://link.aps.org/doi/10.1103/52wh-1z5y}
}

@article{2025ShuiD,
  title = {Ferroelastically Tunable Altermagnets},
  author = {Ding, Ning and Ye, Haoshen and Wang, Shan-Shan and Dong, Shuai},
  year = 2025,
  month = dec,
  journal = {Physical Review B},
  volume = {112},
  number = {22},
  pages = {L220410},
  publisher = {American Physical Society},
  doi = {10.1103/m33v-xwn3},
  urldate = {2026-06-30},
  langid = {english}
}

@article{2025NM-Hall,
  title = {Spontaneous {{Hall}} Effect Induced by Collinear Antiferromagnetic Order at Room Temperature},
  author = {Takagi, Rina and Hirakida, Ryosuke and Settai, Yuki and Oiwa, Rikuto and Takagi, Hirotaka and Kitaori, Aki and Yamauchi, Kensei and Inoue, Hiroki and Yamaura, Jun-ichi and {Nishio-Hamane}, Daisuke and Itoh, Shinichi and Aji, Seno and Saito, Hiraku and Nakajima, Taro and Nomoto, Takuya and Arita, Ryotaro and Seki, Shinichiro},
  year = 2025,
  month = jan,
  journal = {Nature Materials},
  volume = {24},
  number = {1},
  pages = {63--68},
  issn = {1476-1122, 1476-4660},
  doi = {10.1038/s41563-024-02058-w},
  urldate = {2025-12-09},
  langid = {english},
}

@article{2023PRL-spin2charge,
  title = {Efficient {{Spin-to-Charge Conversion}} via {{Altermagnetic Spin Splitting Effect}} in {{Antiferromagnet}} {$<$}math Xmlns="{{http://www.w3.org/1998/Math/MathML"}} Display="inline"{$><$}mrow{$><$}msub{$><$}mrow{$><$}mi{$>$}{{RuO}}{$<$}/Mi{$><$}/Mrow{$><$}mn{$>$}2{$<$}/Mn{$><$}/Msub{$><$}/Mrow{$><$}/Math{$>$}},
  shorttitle = {Efficient {{Spin-to-Charge Conversion}} via {{Altermagnetic Spin Splitting Effect}} in {{Antiferromagnet}}},
  author = {Bai, H. and Zhang, Y. C. and Zhou, Y. J. and Chen, P. and Wan, C. H. and Han, L. and Zhu, W. X. and Liang, S. X. and Su, Y. C. and Han, X. F. and Pan, F. and Song, C.},
  year = 2023,
  month = may,
  journal = {Physical Review Letters},
  volume = {130},
  number = {21},
  pages = {216701},
  publisher = {American Physical Society},
  doi = {10.1103/PhysRevLett.130.216701},
  urldate = {2026-06-30},
  langid = {english}
}

@article{2026PRL-Proximity,
  title = {Altermagnetic Proximity Effect},
  author = {Zhu, Ziye and Huang, Richang and Chen, Xianzhang and Cui, Zhou and Duan, Xunkai and Zhang, Jiayong and \ifmmode \check{Z}\else \v{Z}\fi{}uti\ifmmode \acute{c}\else \'{c}\fi{}, Igor and Zhou, Tong},
  journal = {Phys. Rev. Lett.},
  volume = {136},
  issue = {18},
  pages = {186702},
  numpages = {8},
  year = {2026},
  month = {May},
  publisher = {American Physical Society},
  doi = {10.1103/kqy8-myz1},
  url = {https://link.aps.org/doi/10.1103/kqy8-myz1}
}

@article{2024PRX-Octupoles,
  title = {Ferroically {{Ordered Magnetic Octupoles}} in d -{{Wave Altermagnets}}},
  author = {Bhowal, Sayantika and Spaldin, Nicola A.},
  year = 2024,
  month = feb,
  journal = {Physical Review X},
  volume = {14},
  number = {1},
  pages = {011019},
  issn = {2160-3308},
  doi = {10.1103/PhysRevX.14.011019},
  urldate = {2025-12-09},
  langid = {english},
}

@article{2024NC-SBZ,
  title = {Finite-Momentum {{Cooper}} Pairing in Proximitized Altermagnets},
  author = {Zhang, Song-Bo and Hu, Lun-Hui and Neupert, Titus},
  year = 2024,
  month = feb,
  journal = {Nature Communications},
  volume = {15},
  number = {1},
  pages = {1801},
  issn = {2041-1723},
  doi = {10.1038/s41467-024-45951-3},
  urldate = {2024-03-18},
  langid = {english}
}

@article{2016Science-Electrical,
doi = {10.1126/science.aab1031},
author = {P. Wadley  and B. Howells  and J. Železný  and C. Andrews  and V. Hills  and R. P. Campion  and V. Novák  and K. Olejník  and F. Maccherozzi  and S. S. Dhesi  and S. Y. Martin  and T. Wagner  and J. Wunderlich  and F. Freimuth  and Y. Mokrousov  and J. Kuneš  and J. S. Chauhan  and M. J. Grzybowski  and A. W. Rushforth  and K. W. Edmonds  and B. L. Gallagher  and T. Jungwirth },
title = {Electrical switching of an antiferromagnet},
journal = {Science},
volume = {351},
number = {6273},
pages = {587-590},
year = {2016},
doi = {10.1126/science.aab1031},
URL = {https://www.science.org/doi/abs/10.1126/science.aab1031}
}

@article{2026RuiCX,
  title = {Anomalous-{{Hall N\'eel}} Textures in Altermagnetic Materials},
  author = {Xiao, Rui-Chun and Li, Hui and Han, Hui and Gan, Wei and Yang, Mengmeng and Shao, Ding-Fu and Zhang, Shu-Hui and Gao, Yang and Tian, Mingliang and Zhou, Jianhui},
  year = 2026,
  month = jan,
  journal = {Science China Physics, Mechanics \& Astronomy},
  volume = {69},
  number = {1},
  pages = {217511},
  publisher = {Science China Press},
  issn = {1869-1927},
  doi = {10.1007/s11433-025-2769-6},
  urldate = {2026-06-30},
  copyright = {2025 Science China Press},
  langid = {english}
}

@article{2024Science-180,
author = {Lei Han  and Xizhi Fu  and Rui Peng  and Xingkai Cheng  and Jiankun Dai  and Liangyang Liu  and Yidian Li  and Yichi Zhang  and Wenxuan Zhu  and Hua Bai  and Yongjian Zhou  and Shixuan Liang  and Chong Chen  and Qian Wang  and Xianzhe Chen  and Luyi Yang  and Yang Zhang  and Cheng Song  and Junwei Liu  and Feng Pan },
title = {Electrical 180° switching of Néel vector in spin-splitting antiferromagnet},
journal = {Science Advances},
volume = {10},
number = {4},
pages = {eadn0479},
year = {2024},
doi = {10.1126/sciadv.adn0479}
}

@article{2018NC-DetNeel,
  title = {Electrically Induced and Detected {{N\'eel}} Vector Reversal in a Collinear Antiferromagnet},
  author = {Godinho, J. and Reichlov{\'a}, H. and Kriegner, D. and Nov{\'a}k, V. and Olejn{\'i}k, K. and Ka{\v s}par, Z. and {\v S}ob{\'a}{\v n}, Z. and Wadley, P. and Campion, R. P. and Otxoa, R. M. and Roy, P. E. and {\v Z}elezn{\'y}, J. and Jungwirth, T. and Wunderlich, J.},
  year = 2018,
  month = nov,
  journal = {Nature Communications},
  volume = {9},
  number = {1},
  pages = {4686},
  publisher = {Nature Publishing Group},
  issn = {2041-1723},
  doi = {10.1038/s41467-018-07092-2},
  urldate = {2026-07-03},
  copyright = {2018 The Author(s)},
  langid = {english}
}

@article{2017NPho-neel,
  title = {Optical Determination of the {{N\'eel}} Vector in a {{CuMnAs}} Thin-Film Antiferromagnet},
  author = {Saidl, V. and N{\v e}mec, P. and Wadley, P. and Hills, V. and Campion, R. P. and Nov{\'a}k, V. and Edmonds, K. W. and Maccherozzi, F. and Dhesi, S. S. and Gallagher, B. L. and Troj{\'a}nek, F. and Kune{\v s}, J. and {\v Z}elezn{\'y}, J. and Mal{\'y}, P. and Jungwirth, T.},
  year = 2017,
  month = feb,
  journal = {Nature Photonics},
  volume = {11},
  number = {2},
  pages = {91--96},
  publisher = {Nature Publishing Group},
  issn = {1749-4893},
  doi = {10.1038/nphoton.2016.255},
  urldate = {2026-07-03},
  copyright = {2016 Springer Nature Limited},
  langid = {english}
}

@article{2025Probing,
  title = {Probing k-Space Alternating Spin Polarization via the Anomalous Hall Effect},
  shorttitle = {Probing},
  author = {Chen, Rui and Wang, Zi-Ming and Wu, Ke and Sun, Hai-Peng and Zhou, Bin and Wang, Rui and Xu, Dong-Hui},
  year = 2025,
  month = aug,
  journal = {Physical Review Letters},
  volume = {135},
  number = {9},
  pages = {096602},
  publisher = {American Physical Society},
  doi = {10.1103/yrs7-m6zy},
  urldate = {2026-06-30},
  langid = {english}
}

@article{2026PRL-Photocurrent,
  title = {Band-Geometry-Driven Spin Photocurrent in Centrosymmetric Altermagnets},
  author = {Dong, Ruizhi and Xiao, Yihua and Fei, Ruixiang},
  journal = {Phys. Rev. Lett.},
  volume = {136},
  issue = {21},
  pages = {216702},
  numpages = {7},
  year = {2026},
  month = {May},
  publisher = {American Physical Society},
  doi = {10.1103/g85x-rgxm},
  url = {https://link.aps.org/doi/10.1103/g85x-rgxm}
}

@article{1984Berry,
  title = {Quantal Phase Factors Accompanying Adiabatic Changes},
  author = {Berry, Michael Victor},
  year = 1984,
  month = mar,
  journal = {Proceedings of the Royal Society of London. A. Mathematical and Physical Sciences},
  volume = {392},
  number = {1802},
  pages = {45--57},
  publisher = {The Royal Society},
  issn = {0080-4630},
  doi = {10.1098/rspa.1984.0023},
  urldate = {2026-06-24},
  langid = {english}
}

@article{2024-QGTAlter,
  title = {Quantum Geometry Induced Nonlinear Transport in Altermagnets},
  author = {Fang, Yuan and Cano, Jennifer and Ghorashi, Sayed Ali Akbar},
  journal = {Phys. Rev. Lett.},
  volume = {133},
  issue = {10},
  pages = {106701},
  numpages = {7},
  year = {2024},
  month = {Sep},
  publisher = {American Physical Society},
  doi = {10.1103/PhysRevLett.133.106701}
}

@article{2014PRL-Angular,
  title = {Angular {{Momentum}} of {{Phonons}} and the {{Einstein}}--de {{Haas Effect}}},
  author = {Zhang, Lifa and Niu, Qian},
  year = 2014,
  month = feb,
  journal = {Physical Review Letters},
  volume = {112},
  number = {8},
  pages = {085503},
  publisher = {American Physical Society},
  doi = {10.1103/PhysRevLett.112.085503},
  urldate = {2026-06-24},
  langid = {english}
  }

@article{2025NP-ChiPhonon,
  title = {Chiral Phonons},
  author = {Juraschek, Dominik M. and Geilhufe, R. Matthias and Zhu, Hanyu and Basini, Martina and Baum, Peter and Baydin, Andrey and Chaudhary, Swati and Fechner, Michael and Flebus, Benedetta and Grissonnanche, Gael and Kirilyuk, Andrei I. and Lemeshko, Mikhail and Maehrlein, Sebastian F. and Mignolet, Maxime and Murakami, Shuichi and Niu, Qian and Nowak, Ulrich and Romao, Carl P. and Rostami, Habib and Satoh, Takuya and Spaldin, Nicola A. and Ueda, Hiroki and Zhang, Lifa},
  year = 2025,
  month = oct,
  journal = {Nature Physics},
  volume = {21},
  number = {10},
  pages = {1532--1540},
  issn = {1745-2473, 1745-2481},
  doi = {10.1038/s41567-025-03001-9},
  urldate = {2026-09-16},
  langid = {english},
}

@article{2017PRL-Circular,
  title = {Circular {{Phonon Dichroism}} in {{Weyl Semimetals}}},
  author = {Liu, Donghao and Shi, Junren},
  year = 2017,
  month = aug,
  journal = {Physical Review Letters},
  volume = {119},
  number = {7},
  pages = {075301},
  publisher = {American Physical Society},
  doi = {10.1103/PhysRevLett.119.075301},
  urldate = {2026-06-24},
  langid = {english}
}

@article{2022PRB-Anomalous,
  title = {Anomalous Circular Phonon Dichroism in Transition Metal Dichalcogenides},
  author = {Shan, Wen-Yu},
  year = 2022,
  month = mar,
  journal = {Physical Review B},
  volume = {105},
  number = {12},
  pages = {L121302},
  publisher = {American Physical Society},
  doi = {10.1103/PhysRevB.105.L121302},
  urldate = {2026-06-24},
  langid = {english}
}

@article{2025arxiv-phonon,
  title={Phonon dichroisms revealing unusual electronic quantum geometry},
  author={Li, Ding and Yang, Guoao and Qin, Tao and Zhou, Jianhui and Yao, Yugui},
  journal={arXiv preprint arXiv:2511.16141},
  year={2025}
}

@article{2024PRB-EzawaNon,
  title = {Intrinsic nonlinear conductivity induced by quantum geometry in altermagnets and measurement of the in-plane N\'eel vector},
  author = {Ezawa, Motohiko},
  journal = {Phys. Rev. B},
  volume = {110},
  issue = {24},
  pages = {L241405},
  numpages = {6},
  year = {2024},
  month = {Dec},
  publisher = {American Physical Society},
  doi = {10.1103/PhysRevB.110.L241405}
}

@article{2025PRB-MCDAlter,
  title = {N\'eel vector dependence of x-ray magnetic circular dichroism in altermagnets},
  author = {Kune\ifmmode \check{s}\else \v{s}\fi{}, J.},
  journal = {Phys. Rev. B},
  volume = {112},
  issue = {18},
  pages = {184415},
  numpages = {6},
  year = {2025},
  month = {Nov},
  publisher = {American Physical Society},
  doi = {10.1103/s1rn-qw4h}
}

@article{2010RMP-niuqian,
  title = {Berry Phase Effects on Electronic Properties},
  author = {Xiao, Di and Chang, Ming-Che and Niu, Qian},
  year = 2010,
  month = jul,
  journal = {Reviews of Modern Physics},
  volume = {82},
  number = {3},
  pages = {1959},
  publisher = {American Physical Society},
  doi = {10.1103/RevModPhys.82.1959},
  urldate = {2026-06-24},
  langid = {english}
}

@article{2009Manybody,
  title={Fundamentals of Many-body Physics},
  author={ Nolting, Wolfgang  and  Brewer, William D.  and translator},
  journal={Springer Berlin Heidelberg},
  year={2009},
}

@article{2023PRR-Nonreciprocal,
  title = {Nonreciprocal Phonon Dichroism Induced by {{Fermi}} Pocket Anisotropy in Two-Dimensional {{Dirac}} Materials},
  author = {Shan, Wen-Yu},
  year = 2023,
  month = may,
  journal = {Physical Review Research},
  volume = {5},
  number = {2},
  pages = {L022038},
  publisher = {American Physical Society},
  doi = {10.1103/PhysRevResearch.5.L022038},
  urldate = {2026-06-24},
  langid = {english}}

@article{2012PRL-space,
  title = {Space {{Dependent Fermi Velocity}} in {{Strained Graphene}}},
  author = {de Juan, Fernando and Sturla, Mauricio and Vozmediano, Mar{\'i}a A. H.},
  year = 2012,
  month = may,
  journal = {Physical Review Letters},
  volume = {108},
  number = {22},
  pages = {227205},
  publisher = {American Physical Society},
  doi = {10.1103/PhysRevLett.108.227205},
  urldate = {2026-06-24},
  langid = {english}
}

@article{2013PRB-Gauge,
  title = {Gauge Fields from Strain in Graphene},
  author = {de Juan, Fernando and Ma{\~n}es, Juan L. and Vozmediano, Mar{\'i}a A. H.},
  year = 2013,
  month = apr,
  journal = {Physical Review B},
  volume = {87},
  number = {16},
  pages = {165131},
  publisher = {American Physical Society},
  doi = {10.1103/PhysRevB.87.165131},
  urldate = {2026-06-24},
  langid = {english}
}

@article{2009PRL-strain,
  title = {Strain {{Engineering}} of {{Graphene}}'s {{Electronic Structure}}},
  author = {Pereira, Vitor M. and Neto, A. H. Castro},
  year = 2009,
  month = jul,
  journal = {Physical Review Letters},
  volume = {103},
  number = {4},
  pages = {046801},
  publisher = {American Physical Society},
  doi = {10.1103/PhysRevLett.103.046801},
  urldate = {2026-06-24},
  langid = {english}
  }

@article{2007PRB-Symmetry,
  title = {Symmetry-Based Approach to Electron-Phonon Interactions in Graphene},
  author = {Ma{\~n}es, J. L.},
  year = 2007,
  month = jul,
  journal = {Physical Review B},
  volume = {76},
  number = {4},
  pages = {045430},
  publisher = {American Physical Society},
  doi = {10.1103/PhysRevB.76.045430},
  urldate = {2026-06-24},
  langid = {english}
}

@article{1966PR-SYM,
  title = {Space-{{Time Symmetry}} of {{Transport Coefficients}}},
  author = {Kleiner, W. H.},
  year = 1966,
  month = feb,
  journal = {Physical Review},
  volume = {142},
  number = {2},
  pages = {318--326},
  issn = {0031-899X},
  doi = {10.1103/PhysRev.142.318},
  urldate = {2023-01-05},
  langid = {english},
}

@article{2025PRL-YanSun,
  title = {d-Wave Flat Fermi Surface in Altermagnets Enables Maximum Charge-to-Spin Conversion},
  author = {Lai, Junwen and Yu, Tianye and Liu, Peitao and Liu, Long and Xing, Guozhong and Chen, Xing-Qiu and Sun, Yan},
  year = 2025,
  month = dec,
  journal = {Physical Review Letters},
  volume = {135},
  number = {25},
  pages = {256702},
  publisher = {American Physical Society},
  doi = {10.1103/bf1n-sxdl},
  urldate = {2026-06-30},
  langid = {english}
}

@article{2020SciA-Crystal,
doi = {10.1126/sciadv.aaz8809},
author = {Libor Šmejkal  and Rafael González-Hernández  and T. Jungwirth  and J. Sinova },
title = {Crystal time-reversal symmetry breaking and spontaneous Hall effect in collinear antiferromagnets},
journal = {Science Advances},
volume = {6},
number = {23},
pages = {eaaz8809},
year = {2020},
doi = {10.1126/sciadv.aaz8809},
URL = {https://www.science.org/doi/abs/10.1126/sciadv.aaz8809}
}

@article{2024SciAdv-Observation,
  title = {Observation of Time-Reversal Symmetry Breaking in the Band Structure of Altermagnetic {{RuO}} {\textsubscript{2}}},
  author = {Fedchenko, Olena and Min{\'a}r, Jan and Akashdeep, Akashdeep and D'Souza, Sunil Wilfred and Vasilyev, Dmitry and Tkach, Olena and Odenbreit, Lukas and Nguyen, Quynh and Kutnyakhov, Dmytro and Wind, Nils and Wenthaus, Lukas and Scholz, Markus and Rossnagel, Kai and Hoesch, Moritz and Aeschlimann, Martin and Stadtm{\"u}ller, Benjamin and Kl{\"a}ui, Mathias and Sch{\"o}nhense, Gerd and Jungwirth, Tomas and Hellenes, Anna Birk and Jakob, Gerhard and {\v S}mejkal, Libor and Sinova, Jairo and Elmers, Hans-Joachim},
  year = 2024,
  month = feb,
  journal = {Science Advances},
  volume = {10},
  number = {5},
  pages = {eadj4883},
  issn = {2375-2548},
  doi = {10.1126/sciadv.adj4883},
  urldate = {2024-07-12},
  langid = {english},
}

@article{2026PRL-CCL,
  title = {Antiferroaxial {{Altermagnetism}}},
  author = {Liu, Yichen and Liu, Cheng-Cheng},
  year = 2026,
  month = jun,
  journal = {Physical Review Letters},
  volume = {136},
  number = {25},
  pages = {256709},
  issn = {0031-9007, 1079-7114},
  doi = {10.1103/21z4-c9p2},
  urldate = {2026-06-30},
  langid = {english}
}

\end{document}